\documentclass{iaus}
\usepackage{graphicx}
\usepackage{natbib}

\begin{document}

\title[Spitzer/IRS Spectroscopy of Seyferts] 
{Mid-IR properties of Seyferts: Spitzer/IRS spectroscopy of the IRAS 12$\mu$m Seyfert sample}

\author[V. Charmandaris et al.]   
{Vassilis Charmandaris$^{1,2}$, Yanling Wu$^3$, Jiasheng Huang$^4$, Luigi Spinoglio$^{5,6}$
 \and
Silvia Tommasin$^{5,6}$}

\affiliation{$^1$University of Crete, Department of Physics, GR-71003,
  Heraklion, Greece\\ 
email: {\tt vassilis@physics.uoc.gr} \\[\affilskip]
$^2$IESL/Foundation for Research and Technology - Hellas,
  GR-71110, Heraklion, Greece, and Chercheur Associ\'e, Observatoire
  de Paris, F-75014, Paris, France \\[\affilskip]
$^3$Infrared Processing and Analysis Center, Caltech,  Pasadena, CA 91125, USA \\[\affilskip]
$^4$Harvard-Smithsonian Center for Astrophysics, Cambridge, MA, 02138, USA \\[\affilskip]
$^5$Istituto di Fisica dello Spazio Interplanetario,
  INAF, I-00133 Rome, Italy \\[\affilskip]
$^6$Dipartimento di Fisica, Universita di Roma, La Sapienza, Rome,
  Italy \\[\affilskip]
}

\pubyear{2009}
\volume{267}  
\pagerange{1--126}
\setcounter{page}{1}
\jname{Co-Evolution of Central Black Holes and Galaxies}
\editors{B.M.\ Peterson, R.S.\ Somerville, \& T.\ Storchi-Bergmann, eds.}

\maketitle
\vspace*{-0.2cm}

\begin{abstract}
  We performed an analysis of the mid-infared properties of the
  12\,$\mu$m Seyfert sample, a complete unbiased 12\,$\mu$m flux
  limited sample of local Seyfert galaxies selected from the {\em
    IRAS} Faint Source Catalog, based on low resolution spectra obtained with the
  Infrared Spectrograph (IRS) on-board {\it Spitzer} Space
  Telescope. A detailed presentation of this analysis is dicussed in Wu et al. (2009). \\
  We find that on average, the 15-30\,$\mu$m slope of the continuum is
  $<$$\alpha_{15-30}$$>$=-0.85$\pm$0.61 for Seyfert 1s and -1.53$\pm$0.84
  for Seyfert 2s, and there is substantial scatter in each
  type. Moreover, nearly 32\% of Seyfert 1s, and 9\% of Seyfert 2s,
  display a peak in the mid-infrared spectrum at 20\,$\mu$m, which is
  attributed to an additional hot dust component.  The Polycyclic
  Aromatic Hydrocarbon (PAH) equivalent width decreases with
  increasing dust temperature, as indicated by the global infrared
  color of the host galaxies. However, no statistical difference in
  PAH equivalent width is detected between the two Seyfert types, 1
  and 2, of the same bolometric luminosity.  Finally, we propose a new
  method to estimate the AGN contribution to the integrated 12\,$\mu$m
  galaxy emission, by subtracting the "star formation" component in
  the Seyfert galaxies, making use of the tight correlation between
  PAH 11.2\,$\mu$m luminosity and 12\,$\mu$m luminosity for star
  forming galaxies.

\keywords{galaxies: active, galaxies: Seyfert, galaxies: nuclei, infrared: galaxies}
\end{abstract}

\firstsection 
\vspace*{-0.2cm}

\section{Introduction}

Active galaxies emit radiation from their nucleus which is due to
accretion onto a super-massive black hole (SMBH) located at the
center. The fraction of the energy emitted from these active nuclei
(AGN), compared with the total bolometric emission of the host can
range from a few percent in moderated luminosity systems (L$_{\rm bol}
<10^{11}$L$_{\odot}$), to more than 90\% in quasars (L$_{\rm bol}
>10^{12}$L$_{\odot}$) \citep[see][ and references therein]{Ho08}. As a
subclass, Seyfert galaxies are the nearest and brightest AGNs, with
2-10keV X-ray luminosities less than $\sim10^{44}$ergs$^{-1}$ and
their observed spectral line emission originates principally from
highly ionized gas. Seyferts have been studied at many wavelengths and
classified as Seyfert 1s (Sy 1s) and Seyfert 2s (Sy 2s), with the type
1s displaying features of both broad (FWHM$>$2000 km s$^{-1}$) and
narrow emission lines, while the type 2s only narrow-line emission.

The differences between the two Seyfert types have been an intense
field of study for many years. Are they due to intrinsic differences
in their physical properties, or are they simply a result of dust
obscuration that hides the broad-line region in Sy 2s? A so-called
unified model has been proposed \citep[see][]{Antonucci93, Urry95},
suggesting that Sy 1s and Sy 2s are essentially the same objects
viewed at different angles. A dust torus surrounding the central
engine blocks the optical light when viewed edge on (Sy 2s) and allows
the nucleus to be seen when viewed face on (Sy 1s). Optical spectra in
polarized light \citep{Antonucci85} have indeed demonstrated for
several Sy 2s the presence of broad lines, confirming for these
objects the validity of the unified model.  However, the exact nature
of this orientation-dependent obscuration is not clear yet. Recently,
more elaborate models, notably the ones of \citet{Elitzur08},
\citet{Nenkova08}, and \citet{Thompson09} suggest that the same
observational constraints can also be explained with discrete dense
molecular clouds, without the need of a torus geometry.

Mid-IR spectroscopy is a powerful tool to examine the nature of the
emission from AGNs, as well as the nuclear star-formation
activity. Since IR observations are much less affected by dust
extinction than those at shorter wavelengths, they have been
instrumental in the study of obscured emission from optically thick
regions in AGNs. This is crucial to understand the physical process of
galaxy evolution. With the advent of the {\em Infrared Space
  Observatory (ISO)}, local Seyferts have been studied by several
groups \citep[see][ for a review]{Verma05}. The recent launch of the
{\em Spitzer} Space Telescope \citep{Werner04} has enabled the study
of large samples of AGN with substantially better sensitivity and
spatial resolution, in an effort to quantify their mid-IR properties
\citep{Buchanan06, Sturm06, Deo07, Gorjian07, Hao07, Tommasin08}.

We have used archival Spitzer/IRS \citep{Houck04} observation and have
embarked in a detailed study Seyferts of the extended 12\,$\mu$m
galaxy sample.  This sample of 116 Seyfers appears to be the best for
studying in an unbiased manner the affects of an AGN to physical
properties of a galaxy, since all galaxies emit a nearly constant
fraction ($\sim$7\%) of their bolometric luminosity at 12\,$\mu$m
\citep{Spinoglio89}. A total of 103 Seyferts have been observed by
various Spitzer programs using the low-resolution (R$\sim$64-128)
modules of IRS. Among these objects, 47 are optically classified as Sy
1s and 56 as Sy 2s \citep{Rush93}. The details of this work are
summarized here and can be found in \citet{Wu09}. High-resolution
(R$\sim$600) IRS spectroscopy for 110 objects are also available and
their analysis is presented in \citet{Tommasin08, Tommasin09}.

\vspace*{-0.2cm}
\section{Results}                                        

\subsection{Global Mid-IR spectra of Seyfert Galaxies}

It has been well established that the mid-IR spectra of Seyfert
galaxies display a variety of features \citep[see][and references
therein]{Clavel00, Verma05, Weedman05, Buchanan06, Hao07}. This is
understood since, despite the optical classification of their nuclear
activity, emission from the circumnuclear region, as well as of the
host galaxy, also influences the integrated mid-IR spectrum of the
source.

The IRS spectra for 47 Sy 1s and 54 Sy 2s with full 5.5-35\,$\mu$m
spectral coverage, normalized at the wavelength of 22\,$\mu$m, are
averaged and plotted in Figure \ref{fig:avespect}. For comparison, we
over-plot the average starburst template from \citet{Brandl06}. It is
clear that the mid-IR continuum slope of the average Sy 1 spectrum is
shallower than that of Sy 2, while the starburst template has the
steepest spectral slope, indicating a different mixture of hot/cold
dust component in these galaxies \citep[also see][]{Hao07}. This would
be consistent with the interpretation that our mid-IR spectra of Sy 2s
display a strong starburst contribution, possibly due to circumnuclear
star formation activity included in the aperture we used to extract
the spectra from. PAH emission, which is a good tracer of star
formation activity \citep{Forster04}, can be detected in the average
spectra of both Seyfert types, while it is most prominent in the
average starburst spectrum. PAH emission originates from
photo-dissociation region (PDR) and can easily be destroyed by the
UV/X-ray photons in strong a radiation field produced near massive
stars and/or an accretion disk surrounding a SMBH.

\begin{figure}[!h]
\begin{center}
 \includegraphics[width=5in]{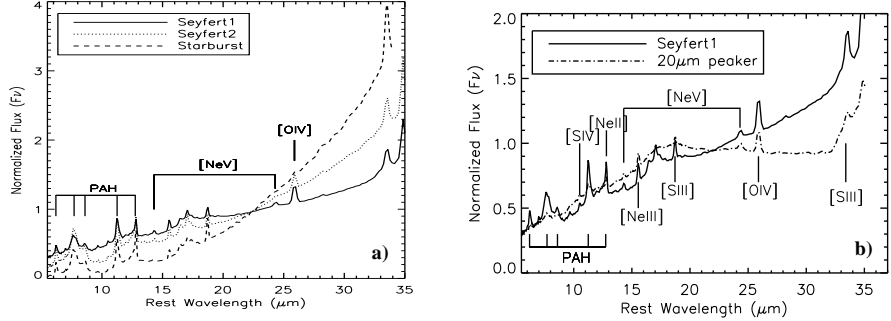}
 \caption{a) A comparison among the average mid-IR spectrum of Sy 1s
   (solid line) and Sy 2s (dotted line) of the 12\,$\mu$m sample, as
   well as the starbursts (dashed line) of \citet{Brandl06}. All
   spectra have been normalized at 22\,$\mu$m. Note that the
   high-ionization fine-structure lines of [OIV]\,25.89\,$\mu$m are
   present in all three spectra, while [NeV]\,14.3/24.3\,$\mu$m are
   only present in the average spectra of the two Seyfert types. b) A
   comparison between the average mid-IR spectrum of ``20\,$\mu$m
   peakers'' (dash-dotted line) and Sy 1s (solid line) of our
   sample. All spectra have been normalized at 22\,$\mu$m.}
  \label{fig:avespect}
\end{center}
\end{figure}

In the 12\,$\mu$m Seyfert sample, we detect PAH emission in 37 Sy 1s
and 53 Sy 2s, that is 78\% and 93\% for each type respectively. This
is expected since the apertures we used to extract the mid-IR spectra
for the 12$\mu$m sample correspond in most cases to areas of more than
1~kpc in linear dimensions. As a result, emission from the PDRs
associated with the extended circumnuclear region and the disk of the
host galaxy is also encompassed within the observed spectrum.  High
ionization fine-structure lines, such as
[NeV]14.32\,$\mu$m/24.32\,$\mu$m, are clearly detected even in the
low-resolution average spectrum of Sy 1. Since [NeV] has an ionization
potential of 97eV serves as an unambiguous indicator of an AGN. This
signature is also visible, though rather weak, in the average spectrum
of Sy 2, while it is absent in the average starburst template . Even
though the low-resolution module of IRS was not designed for studying
fine-structure lines, we are still able to detect [NeV] emission in 29
Sy 1s and 32 Sy 2s, roughly 60\% for both types.  Another high
ionization line, [OIV]\,25.89\,$\mu$m, with ionization potential of
54eV, also appears in both Seyfert types (42 Sy 1s and 41 Sy 2s), and
is stronger in the average spectrum of Sy 1. The [OIV] emission line
can be powered by shocks in intense star forming regions or AGNs
\citep[see][and references therein]{Bernard-Salas09, Hao09}. In our
sample it is probably powered by both, given the large aperture we
adopted for spectral extraction.  More details and a complete analysis
of mid-IR fine-structure lines for 29 galaxies from the 12\,$\mu$m
Seyfert sample are presented in \citet{Tommasin08}, while the work for
the entire sample is in progress (Tommasin et al. 2009).

A thorough examination of the spectra, reveals that 15 Sy 1s and 4 Sy
2s, have a F$_{20}$/F$_{30}$$\ge$0.95. We call these objects
``20\,$\mu$m peakers'' and we display their average spectrum in Figure
1b). In addition to their characteristic continuum shape, a number of
other differences between the ``20\,$\mu$m peakers'' and Sy 1s are
also evident. PAH emission, which is clearly detected in the average
Sy 1 spectrum, appears to be rather weak in the average ``20\,$\mu$m
peaker'' spectrum. The high-ionization lines of [NeV] and [OIV] are
seen in both spectra with similar strength, while low-ionization
lines, especially [NeII] and [SIII], are much weaker in the average
spectrum of ``20\,$\mu$m peakers''. If we calculate the infrared color
of a galaxy using the ratio of F$_{25}$/F$_{60}$, we find an average
value of 0.75 for the ``20\,$\mu$m peakers'', while it is 0.30 for the
other ``non-20\,$\mu$m peaker'' Sy 1s in the 12\,$\mu$m sample.
Finally, the average IR luminosities of the ``20\,$\mu$m peakers'' and
Sy 1s do not show significant difference, with log(L$_{\rm
  IR}/L_\odot$)=10.96 for the former and log(L$_{\rm
  IR}/L_\odot$)=10.86 for the latter. These results are consistent
with the ``20\,$\mu$m peakers'' being AGNs with a dominant hot dust
emission from a small grain population heated to effective
temperatures of $\sim$ 150K and a possible contribution due to the
distinct emissivity of astronomical silicates at 18$\mu$m.  Their
radiation field must also be stronger than a typical Sy 1, since it
destroys the PAH molecules around the nuclear region more
efficiently \citep[see][ for more details]{Wu09}.

\subsection{The PAH emission in the 12$\mu$m Seyferts}

To quantify the strength of PAH emission, we follow the usual approach
and measure the fluxes and equivalent widths (EWs) of the 6.2 and
11.2\,$\mu$m PAH features from their mid-IR spectra.

\begin{figure}[!h]
\begin{center}
 \includegraphics[width=5in]{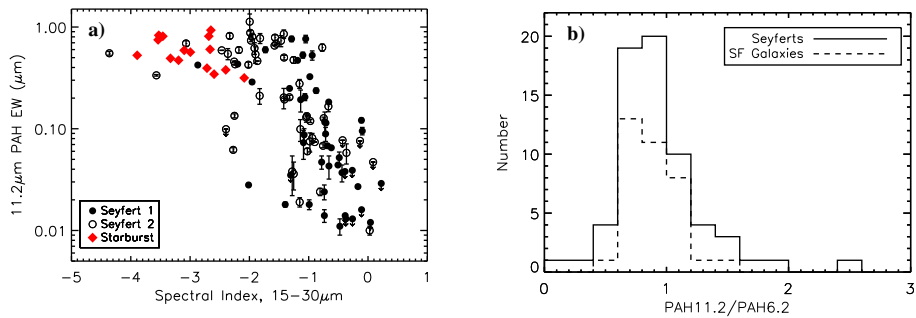}
 \caption{a) The 15-30\,$\mu$m spectral index vs 11.2\,$\mu$m PAH EW for
    the 12\,$\mu$m Seyfert sample.The filled circles are Seyfert 1s,
    the open circles are Seyfert 2s, while the diamonds denote the
    starburst galaxies from \citet{Brandl06}.Note that the PAH EWs of
    Seyferts are progressively suppressed as their 15 to 30\,$\mu$m
    continuum slopes flatten. b) A histogram of the flux ratio of the 11.2\,$\mu$m PAH to
    the 6.2\,$\mu$m PAH feature. The solid line indicates the values
    of the 12\,$\mu$m Seyfert sample while the dashed line indicates
    those of the SF galaxies from the \citet{Brandl06} and
    \citet{Smith07a}. Galaxies with only upper limits measured for the
    aromatic features are excluded from this plot. Both the SF
    galaxies and the Seyferts appear to have similar distribution of
    the 11.2\,$\mu$m/6.2\,$\mu$m PAH flux ratios, indicating that
    globally the chemical structure of the aromatic features observed
    in Seyfert nuclei are likely very similar to those seen in SF
    galaxies.}
  \label{fig:avespect}
\end{center}
\end{figure}

In Figure 2a, we plot the 15-30\,$\mu$m spectral index for the
12\,$\mu$m Seyfert sample as a function of their 11.2\,$\mu$m PAH
EWs. The diamonds indicate the starburst galaxies from
\citet{Brandl06}. A general trend of the PAH EWs decreasing as a
function of 15-30\,$\mu$m spectral index is observed in Figure 2a,
even though it is much weaker than the anti-correlation presented by
\citet{Deo07}.  Starburst galaxies are located on the upper left
corner of the plot, having very steep spectral slopes, with
$<\alpha_{15-30}>$=-3.02$\pm$0.50, and large PAH EWs, nearly
0.7$\mu$m. Seyfert galaxies spread over a considerably larger range in
spectral slopes as well as PAH EWs. Sy 1s and Sy 2s are mixed on the
plot. On average, the 15-30\,$\mu$m spectral index is
$<\alpha_{15-30}>$=-0.85$\pm$0.61 for Sy 1s and
$<\alpha_{15-30}>$=-1.53$\pm$0.84 for Sy 2s. Note that although the
mean spectral slope is slightly steeper for Sy 2s, there is
substantial scatter, as is evident by the uncertainties of the mean
for each types.

It is well known that the flux ratio of different PAH emission bands
is a strong function of PAH size and their ionization state
\citep{Draine01}. The 6.2\,$\mu$m PAH emission is due to C-C
stretching mode and the 11.2\,$\mu$m feature is produced by C-H
out-of-plane bending mode \citep{Draine03}. In Figure 2b, we display a
histogram of the 11.2\,$\mu$m to 6.2\,$\mu$m PAH flux ratios for the
12\,$\mu$m Seyferts. Given the relatively small number of starburst
galaxies in the \citet{Brandl06} sample (16 sources), we also included
20 HII galaxies from the SINGS sample of \citet{Smith07a}, thus
increasing the number of SF galaxies to 36 sources and making its
comparison with the Seyferts more statistically meaningful.  From
Figure 2b, we can see that both the Seyferts and SF galaxies,
indicated by the solid and dashed line respectively, appear to have
very similar distribution of PAH 11.2$\mu$m/6.2$\mu$m band
ratios. This implies that even though the harsh radiation field in
AGNs may destroy a substantial amount of the circumnuclear PAH
molecules, and does so preferentially it likely does not do so over a
large volume. Enough molecules in the circumnuclear regions do remain
intact and as a result, the aromatic features that we observe from
Seyferts are essentially identical to those in SF galaxies. The
relative strength of PAH emission can also be used to examine the
validity of the unified AGN model which attributes the variation in
AGN types as the result of dust obscuration and relative orientation
of the line of sight to the nucleus \citep{Antonucci93}. Sy 1s and Sy
2s are intrinsically the same but appear to be different in the
optical, mainly because of the much larger extinction towards the
nuclear continuum of Sy 2s when viewed edge on.  The latest analysis
of the IRS high-resolution spectra of 87 galaxies from the 12\,$\mu$m
Seyfert sample shows that the average 11.2\,$\mu$m PAH EW is
0.29$\pm$0.38\,$\mu$m for Sy 1s and 0.37$\pm$0.35\,$\mu$m for Sy 2s
(Tommasin et al. 2009, in preparation).

\vspace*{-0.5cm}
\section{What powers the 12\,$\mu$m luminosity in the 12$\mu$m Seyferts?}

Using the data in hand we propose a new statistical method to estimate
the AGN contribution to the total infrared luminosity of a galaxy. The
method, which is presented in detail in \citet[]{Wu09}, relies on the
fact that for starforming galaxies there is a clear correlation
between the L(11.2\,$\mu$m PAH) and L(12\,$\mu$m continnum)
luminosities with an average L(11.2\,$\mu$m PAH)/L(12\,$\mu$m) ratio
of 0.044$\pm$0.010. Since there is no AGN contamination in the
12\,$\mu$m luminosity of star forming galaxies, we can attribute all
mid-IR continuum emission to star formation. Seyfert galaxies do
display a larger scatter in this ratio \citep[see Fig. 14 of
][]{Wu09}. However, we can decompose their 12\,$\mu$m luminosity into
two parts: one contributed by the star formation activity, which is
proportional to their PAH luminosity, and one due to dust heated by
the AGN. If we assume that the star formation component in the
12\,$\mu$m luminosity of Seyferts is associated with the 11.2\,$\mu$m
PAH luminosity in the same manner as in star forming galaxies, then we
can estimate the star formation contribution to the integrated
12\,$\mu$m luminosity of the Seyfert sample. Subtracting this
starforming contribution from the total 12\,$\mu$m luminosity, we can
obtain, in a statistical sense, the corresponding AGN contribution.

\vspace*{-0.2cm}
\section{Conclusions}

Based on the analysis of Spitzer/IRS low resolution spectra for a
complete unbiased sample of Seyfert galaxies selected from the {\em
  IRAS} Faint Source Catalog based on their 12\,$\mu$m fluxes we find
that:

1. The 12\,$\mu$m Seyferts display a variety of mid-IR spectral
shapes. The mid-IR continuum slopes of Sy 1s and Sy 2s are on average
$<\alpha_{15-30}>$=-0.85$\pm$0.61 and -1.53$\pm$0.84 respectively,
though there is substantial scatter for both types. We identify a
group of objects with a local maximum in their mid-IR continuum at
$\sim$20\,$\mu$m, which is likely due to the presence of a warm
$\sim$150 K dust component and 18\,$\mu$m emission from astronomical
silicates. Emission lines, such as the
[NeV]\,14.3\,$\mu$m/24.3\,$\mu$m and [OIV]\,25.9\,$\mu$m lines, known
to be a signature of an AGN are stronger in the average spectra of Sy
1s than those of Sy 2s.

2. PAH emission is detected in both Sy 1s and Sy 2s, with no
statistical difference in the relative strength of PAHs between the
two types. This suggests that the volume responsible for the bulk of
their emission is likely optically thin at $\sim$12\,$\mu$m.

3. The 11.2\,$\mu$m PAH EW of the 12\,$\mu$m Seyfert sample correlates
well with the IRAS color of the galaxies as indicated by the flux
ratio of F$_{25}$/F$_{60}$. PAH emission is more suppressed in warmer
galaxies, in which the strong AGN activity may destroy the PAH
molecules.

4. The FIR luminosities of the 12\,$\mu$m Seyferts are dominated by
star-formation. Their mid-IR luminosity increases by the additional
AGN contribution. A method to estimate the AGN contribution to the
12\,$\mu$m luminosity, in a statistical sense, has been proposed and
applied to the sample.

\end{document}